\documentclass[a4paper,fleqn]{cas-dc}

\usepackage[numbers]{natbib}
\usepackage{subcaption}

\def\tsc#1{\csdef{#1}{\textsc{\lowercase{#1}}\xspace}}
\tsc{WGM}
\tsc{QE}

\begin{document}
\let\WriteBookmarks\relax
\def\floatpagepagefraction{1}
\def\textpagefraction{.001}

\shorttitle{Matching Architecture to Workload}    

\shortauthors{Luca Colombo}  

\title [mode = title]{Matching Architecture to Workload: Comparing RISC-V Parallel Accelerators for Energy-Efficient Embedded Computing}



%

\author[1,2,3]{Luca Colombo}[orcid=0009-0008-4272-1849]

\cormark[1]


\ead{luca.colombo07@universitadipavia.it}


\credit{Conceptualization, Methodology, Investigation, Software, Validation, Data Curation, Formal analysis, Visualization, Writing - original draft}


\affiliation[1]{organization={Fondazione Chips-IT},
            city={Pavia},
            postcode={27100}, 
            state={Lombardia},
            country={Italy}}

\affiliation[2]{organization={Università degli Studi di Pavia},
            city={Pavia},
            postcode={27100}, 
            state={Lombardia},
            country={Italy}}
            
\affiliation[3]{organization={Alma Mater Studiorum},
            city={Bologna},
            postcode={40126}, 
            state={Emilia Romagna},
            country={Italy}}
\affiliation[4]{organization={ETH Zürich},
            city={Zurich},
            postcode={8092}, 
            state={Canton of Zurich},
            country={Switzerland}}
            
\author[1]{Luigi Ghionda}
\credit{Methodology, Software, Validation}
\author[3]{Amirhossein Kiamarzi}
\credit{Software, Validation, Data Curation}
\author[1]{Yvan Tortorella}
\credit{Software, Validation}
\author[1]{Riccardo Giunti}
\credit{Software, Validation}
\author[1]{Niccolò Giuliani}
\credit{Software}
\author[1]{Samuele Colmi}
\credit{Software}
\author[3,4]{Angelo Garofalo}
\credit{Writing – review \& editing, Supervision}
\author[1,3]{Giuseppe Tagliavini}
\credit{Software, Writing – review \& editing}
\author[1,3]{Davide Rossi}
\credit{Project administration, Conceptualization, Methodology, Writing – review \& editing, Supervision}




\cortext[1]{Corresponding author}



\begin{abstract}
Embedded computing demands high performance and energy efficiency across workloads with widely different computational characteristics. Choosing the most suitable execution model therefore requires a quantitative understanding of how each model maps to different workload classes. In this paper, we compare three execution models for energy-efficient embedded computing: scalar multicore, streaming multicore, and vector processing. We use three open-source RISC-V accelerators as representative implementations: the PULP Cluster, the Snitch Cluster, and Spatz, respectively. Their open-source hardware and software stacks enable a controlled comparison based on implementations in the same 12 nm technology and a common benchmark suite. Our results reveal that no single execution model performs best across all workload classes. Spatz delivers the greatest benefits for regular and highly vectorizable compute-intensive applications, achieving up to 4.29× higher performance and 5.17× greater energy efficiency than the other architectures. However, its effectiveness strongly depends on the availability of data-level parallelism and regular computation and memory-access patterns, making it the most sensitive to workload characteristics. The Snitch Cluster performs particularly well on applications with limited parallelism and streamable memory-access patterns, providing up to 5.53× higher performance and 1.45× greater energy efficiency. Conversely, the PULP Cluster is most effective on irregular and control-dominated applications, achieving improvements of up to 5.66× in performance and 2.80× in energy efficiency. Compared with Spatz, however, the greater flexibility offered by PULP and Snitch comes at a non-negligible cost in terms of both area and power.
\end{abstract}





\maketitle%

\section{Introduction}
Modern edge and embedded systems are increasingly tasked with demanding computations that require strict latency guarantees. Workloads once reserved for cloud environments, ranging from linear algebra (LA)~\cite{LINALG} and digital signal processing (DSP)~\cite{10955689} to machine learning (ML)~\cite{8114708} and small or large language models (SLMs/LLMs)~\cite{nguyen-etal-2025-survey}, are now being deployed directly on IoT nodes. However, unlike centralized servers, edge devices face severe limitations in terms of power budget, silicon footprint, and memory bandwidth. Consequently, the primary research challenge has shifted from simply maximizing raw throughput to achieving high energy efficiency while strictly adhering to memory and precision constraints.

A major hurdle in designing efficient edge systems is the sheer diversity of computational patterns these workloads present. Algorithms such as matrix multiplication exhibit highly regular, dense data parallelism, allowing for massive independent operations. Conversely, control-heavy tasks or triangular kernels present irregular, dynamic parallelism that fluctuates over time, making it significantly harder to keep computational units fully utilized. Because the choice of execution model profoundly impacts both resource utilization and energy efficiency, there is a strong motivation to develop heterogeneous architectures where distinct tasks are mapped to the most appropriate compute substrate~\cite{garofalo2025reliabletimepredictableheterogeneoussoc, 9911384, 9560136}.

Historically, based on Flynn’s taxonomy~\cite{5009071}, highly parallel tasks have been tackled using SIMD (Single Instruction, Multiple Data) or SIMT (Single-Instruction, Multiple Thread) models, prominently featured in GPUs. While SIMT architectures excel at processing massive, uniform data arrays~\cite{https://doi.org/10.1111/j.1467-8659.2007.01012.x}, their efficiency degrades sharply in the presence of control-flow divergence. Furthermore, in embedded applications, the mixture of DSP and ML kernels with irregular workloads renders the GPGPUs a suboptimal choice.

In contrast, many embedded parallel platforms adopt a Multiple Instruction Multiple Data (MIMD) execution model. Parallel kernels on such systems are typically expressed using an explicit Single Program Multiple Data (SPMD) programming model, in which the same program is executed by multiple cores operating on distinct partitions of the input data~\cite{DAREMA198811}. In this context, we refer to each core as an execution \textit{thread}, ensuring consistent terminology with other solutions. This approach offers a good balance between programmability and control over work partitioning, data tiling and memory locality, which is critical under the tight memory and bandwidth constraints typical of edge systems.

A representative embodiment of this class is GreenWaves Technologies' GAP9, an ultra-low-power SoC based on the PULP platform with a nine-core RISC-V cluster operating in SPMD fashion~\cite{gap9_greenwaves}. Other commercial systems adopt the same MIMD/SPMD principle at different points of the performance-power envelope, from XMOS' XU316-1024-TQ128, with 16 logical cores distributed over two deterministic tiles~\cite{xmos_xcoreai_datasheet}, to Texas Instruments' TMS320C6678, which scales the same execution model to eight independent C66x DSP cores for higher-throughput signal processing at the cost of area and energy efficiency~\cite{ti_tms320c6678}.

Another execution model, \textit{streaming}, decouples operand generation and movement from computation itself, feeding functional units through dedicated address-generation and streaming hardware rather than explicit load/store instructions in the inner loop, keeping single-issue, area-efficient cores fully utilized even on irregular access patterns. Texas Instruments' C7x DSP exemplifies this paradigm, coupling a VLIW core with programmable Streaming Engines and Address Generators that continuously feed the functional units~\cite{texasinstruments2024c7000}.

Vector processors represent another paradigm in the execution model spectrum. Instead of utilizing a large number of software threads, they expose data parallelism explicitly through vector instructions, amortizing control overhead across multiple lanes. Commercial adoption of this paradigm spans both the Arm~\cite{arm_helium} and RISC-V~\cite{riscv-v-spec} ecosystems. On the Arm side, the Helium M-Profile Vector Extension has been integrated into MCUs such as STMicroelectronics' STM32N6~\cite{st_stm32n6}, Renesas' RA8M1~\cite{renesas_ra8} and Alif Semiconductor's AE302F80C1557LE~\cite{alif_ensemble}. On the RISC-V side, Canaan's Kendryte K230 integrates a T-Head C908 core with a full, ratified RVV 1.0 implementation over 128-bit-wide vector registers~\cite{canaan_k230}.

Crucially, every commercial platform discussed above, regardless of the execution paradigm it embodies, is a closed-source design and its implementation is done in a different or undisclosed technology: micro-architectural details and physical implementation data are undisclosed. This holds true even for the two RISC-V-based examples, GAP9 and K230, which merely expose an open instruction set while the underlying core implementation remains proprietary. This opacity precludes any rigorous, fair PPA comparison across execution paradigms, as neither silicon area, power breakdown, nor synthesis and place-and-route results are accessible.  

RISC-V's open ISA has, however, fostered a growing ecosystem of fully open-source academic implementations that instantiate each of the paradigms discussed so far within a single, consistent tool-chain: PULP Cluster~\cite{7864441} for the MIMD/SPMD model, Snitch Cluster~\cite{9216552} for streaming execution, and Spatz~\cite{10069431} for vector processing. This work leverages exactly this property of the RISC-V ecosystem, implementing and comparing these three designs down to the physical level in the same technology, to provide a fair PPA comparison across the three execution paradigms. Following the choices made by~\cite{10.1145/3801487.3801835} and by
evaluating a broad spectrum of edge workloads, from highly regular kernels to irregular, control-intensive algorithms, this work highlights the intrinsic trade-offs of each execution paradigm. Our findings provide critical insights for the future design of energy-efficient, heterogeneous embedded platforms. The main contributions of this paper are as follows:
\begin{itemize}
    \item A comprehensive cross-domain benchmark suite comprising twenty-two fully profiled kernels from machine learning, digital signal processing, large language models, and linear algebra, implemented and optimized for all three architectures;
    \item A silicon-realistic PPA comparison based on resource-equivalent configurations and complete physical implementations of all three architectures in the same 12\,nm technology, using a common target operating frequency and consistent implementation constraints to isolate the impact of the execution model on performance, power, and area;
    \item A quantitative characterization of how workload properties, including data-level parallelism, control-flow behavior, and memory-access regularity, affect each execution model, identifying the workload classes for which scalar multicore, streaming multicore, and vector architectures provide the highest performance and energy efficiency.
\end{itemize}
Our study demonstrates that compute-unit utilization strongly depends on workload characteristics. For compute-intensive, highly parallel workloads, the Spatz vector cluster outperforms the other architectures by up to 4.29$\times$ in performance and 5.17$\times$ in energy efficiency. The Snitch Cluster is most effective on workloads with limited parallelism and streamable memory-access patterns, achieving up to a 5.53$\times$ speedup and a 1.45$\times$ improvement in energy efficiency. Conversely, the multicore PULP Cluster excels in control-dominated tasks, with performance and energy-efficiency advantages of up to 5.66$\times$ and 2.80$\times$, respectively. While vector execution offers superior efficiency for dense data parallelism, its specialization makes it more sensitive to application structure. Snitch and PULP support a broader range of workload characteristics, but this greater flexibility entails non-negligible area and power overheads compared with the specialized vector architecture.

\section{Related Work}

\begin{table*}[t]
\centering
\resizebox{\textwidth}{!}{%
\begin{tabular}{@{}l ccc c cccc c ccc@{}}
\toprule
 & \multicolumn{8}{c}{Commercial} & & \multicolumn{3}{c}{Open-source (this work)} \\
\cmidrule(lr){2-9}\cmidrule(lr){11-13}
Aspect & GAP9$^\dagger$ & XU316 & TMS320C6678 & TDA4VM & STM32N6 & RA8M1 & AE302F80 & K230$^\dagger$ & & PULP Cluster& Snitch Cluster & Spatz \\
\midrule
ISA                    & RISC-V         & Proprietary    & Proprietary & Proprietary & Armv8.1-M   & Armv8.1-M & Armv8.1-M  & RISC-V      & & RISC-V & RISC-V & RISC-V \\
Number of Cores        & 9+1 FC         & 16             & 8           & 11$^\ddagger$ & 1         & 1   &  2     & 2           & & 8      & 8+1    & 1     \\
Max Frequency          & 240-400\,MHz  & 600-800\,MHz  & 1.4\,GHz    & 1.0\,GHz    & 800\,MHz    & 480\,MHz & 160-400\,MHz & 1.6\,GHz    & & 810\,MHz$^\ast$ & 827\,MHz$^\ast$ & 846\,MHz$^\ast$ \\
Technology Node        & 22\,nm         & $\times$       & 40\,nm      & 16\,nm      & 16\,nm      & 40\,nm  &  $\times$ & $\times$    & & 12\,nm$^\star$ & 12\,nm$^\star$ & 12\,nm$^\star$ \\
\midrule
Scalar/MIMD Execution  & \checkmark     & \checkmark     & \checkmark  & $\times$    & $\times$    & $\times$ &  $\times$ & $\times$    & & \checkmark & $\times$   & $\times$   \\
Streaming Execution    & $\times$       & $\times$       & $\times$    & \checkmark  & $\times$    & $\times$  & $\times$ & $\times$    & & $\times$   & \checkmark & $\times$   \\
Vector Execution       & $\times$       & $\times$       & $\times$    & $\times$    & \checkmark  & \checkmark &\checkmark & \checkmark  & & $\times$   & $\times$   & \checkmark \\
\midrule
Open-Source RTL        & $\times$       & $\times$       & $\times$    & $\times$    & $\times$    & $\times$  & $\times$ & $\times$    & & \checkmark & \checkmark & \checkmark \\
PPA Data Available     & $\times$       & $\times$       & $\times$    & $\times$    & $\times$    & $\times$ & $\times$  & $\times$    & & \checkmark & \checkmark & \checkmark \\
\bottomrule
\end{tabular}%
}
\caption{Comparison of representative commercial systems for each execution paradigm, excluding GPGPUs, and of the three open-source clusters studied in this work. Technology node marked $\times$ is not publicly disclosed by the vendor. RTL and physical-implementation data are unavailable for all commercial designs. $^\dagger$ISA is open-source, but the core implementation remains proprietary. $^\ddagger$TDA4VM includes 11 heterogeneous cores; only one of them is the C7x DSP streaming architecture $^\ast$ Obtained by topographical synthesis, not silicon measurements. $^\star$Technology node used in this work, may differ from the one used in the referenced works.}
\label{tab:Table_1}
\end{table*}
The problem of sustaining high functional-unit utilization across workloads with heterogeneous degrees of data-level parallelism has been addressed through several competing architectural paradigms: Single Instruction Multiple Thread (SIMT) based GPUs, general-purpose superscalar and many-core CPUs, streaming/dataflow accelerators and Cray-style vector machines. The three open-source RISC-V clusters compared in this work are each representative of a different point in this design space. This section reviews each paradigm in turn and situates PULP, Snitch, and Spatz within it. Table~\ref{tab:Table_1} reports the specifics of the works, excluding GPUs, cited in this section.

\subsection{GPGPUs}
General-Purpose computation on GPUs (GPGPUs) utilizes a large number of arithmetic units with many concurrent hardware threads, hiding memory-access latency through dynamic warp scheduling rather than through explicit data reuse. 

Recent micro-architectural studies of NVIDIA's Ampere and Hopper generations~\cite{luo2025dissectingnvidiahopperarchitecture} show that successive GPU generations trade a comparatively modest energy-efficiency improvement for substantial control and operand-staging overhead. Open-source RISC-V GPGPU research is centered on the SIMT execution model, from the FPGA-oriented Vortex~\cite{Vortex} and its cluster-level matrix-unit extension Virgo~\cite{Kim_2025}, to the low-power, TinyAI-oriented e-GPU~\cite{machetti2026egpuopensourceconfigurableriscv}. 

While GPGPUs have gained significant traction thanks to scientific computing and AI inference, the most widely adopted designs are closed source, and, to the best of our knowledge, open-source RISC-V GPGPUs do not provide a complete, publicly reproducible physical-implementation flow, which is required for a reproducible PPA analysis.

\subsection{Scalar Multi-Cores}
Another strategy for sustaining throughput under a tight power and area budget is to replicate many simple, single-issue scalar cores rather than widening the datapath of a single core with SIMD or SIMT lanes. Because each core independently fetches, decodes, and issues its own instruction stream, this MIMD organization tolerates irregular control flow and data-dependent branches far more gracefully than SIMD/SIMT machines, at the cost of replicating front-end logic across cores instead of amortizing it over wide lanes or many hardware threads. Parallel kernels on such systems are typically expressed through the SPMD model, in which every core runs the same program over a distinct partition of the data, with cores tied together by a shared, low-latency L1 memory reachable through a logarithmic or crossbar interconnect. 

Commercial systems adopting this philosophy span the full performance-power envelope: GreenWaves Technologies' GAP9 integrates a single-core fabric controller and a nine-core RISC-V cluster operating in SPMD fashion for near-sensor DSP and ML workloads~\cite{gap9_greenwaves}; XMOS' XU316-1024-TQ128 distributes 16 real-time logical cores over two deterministic tiles, favoring timing predictability over the pipeline complexity of general-purpose multi-cores~\cite{xmos_xcoreai_datasheet}; at the opposite end of the spectrum, Texas Instruments' TMS320C6678 KeyStone DSP scales the same execution model to eight independent, floating-point-capable C66x cores for higher-throughput signal processing~\cite{ti_tms320c6678}. 

None of these systems, however,  disclose implementation data, precluding a reproducible  PPA comparison, as can be seen in Table~\ref{tab:Table_1}. Among open-source designs, this many-core philosophy has proven flexible and adequate for most workloads; in this study we therefore choose the PULP Cluster~\cite{7864441} as the representative of this category, owing to its maturity and its tunable number of cores, which allows a fair, scalable comparison.

\subsection{Streaming/Dataflow Architectures}
The streaming/dataflow paradigm targets the same irregular, memory-bound kernels that challenge SIMT and scalar many-core machines, but pursues efficiency along an orthogonal axis: rather than replicating many independent instruction streams, it decouples data movement from computation by letting dedicated address-generation hardware autonomously stream operands into and out of a single-issue core's datapath, removing the fetch/decode/issue overhead of explicit load and store instructions from the innermost loops. 

Texas Instruments' C7x DSP~\cite{texasinstruments2024c7000} is a representative commercial embodiment of this paradigm, coupling a VLIW core with programmable Streaming Engines and Address Generators that continuously feed wide SIMD functional units, reaching up to 80 GFLOPS/256 GOPS when integrated in automotive-grade SoCs such as the TDA4VM~\cite{ti_tda4vm} (Table~\ref{tab:Table_1}). As with the multi-core designs discussed above, the C7x's Streaming Engines, Address Generators, and physical implementation data are entirely undisclosed, making it unsuitable for a reproducible PPA study. 

In this context, the Snitch Cluster~\cite{9216552}, which augments a compact RV32 scalar core with Stream Semantic Registers (SSRs) binding memory streams directly to floating-point register operands, and a FREP sequencer decoupling loop control from the floating-point datapath, is the ideal open-source representative of the streaming paradigm for the comparison carried out in this work.

\subsection{Vector Processors}
Cray-style vector architectures amortize the fetch, decode, and control overhead of a single instruction across a long vector of data elements processed by multiple parallel lanes, rather than distributing that overhead across independent cores as in the MIMD paradigm, or hiding it behind autonomously generated operand streams as in the streaming paradigm above. This regime is particularly efficient on highly regular, dense data-parallel kernels, at the cost of degrading sharply in the presence of control-flow divergence or irregular memory access patterns. 

Thanks to Arm providing its M-Profile Vector Extension (Helium)
~\cite{arm_helium} and RISC-V standardizing the ''V'' extension (RVV 1.0)~\cite{riscv-v-spec}, commercial adoption of this paradigm has started to gain traction. General-purpose MCUs such as STMicroelectronics' STM32N6, built around a Cortex-M55 core coupled with a dedicated Neural-ART accelerator~\cite{st_stm32n6}, Renesas' RA8M1 based on the Cortex-M85~\cite{renesas_ra8}, and Alif Semiconductor's AE302F80C1557LE, which combines one or two Cortex-M55 cores with Cortex-A32 application cores in a single processor~\cite{alif_ensemble}, are representative of ARM vector processors. 

On to the RISC-V side, Canaan's Kendryte K230 integrates a T-Head C908 core with a full, ratified RVV 1.0 implementation over 128-bit-wide vector registers~\cite{canaan_k230}.

As with the multi-core and streaming platforms above, all four designs are closed source at the implementation level, depicted in Table~\ref{tab:Table_1}. Additionally, the Arm-based MCUs rely on a proprietary ISA, while K230's RISC-V core, though built on an open instruction set, exposes no public RTL or PPA data. Spatz~\cite{10069431} closes this gap as a fully open-source Cray-style vector processor, clustering compact vector units around a shared-TCDM interconnect; the possibility to tune the number of Spatz cores, FPUs, and datapath size made this design the ideal candidate for a fair PPA comparison.

\begin{figure*}[]
	\centering
	\includegraphics[width=1\textwidth]{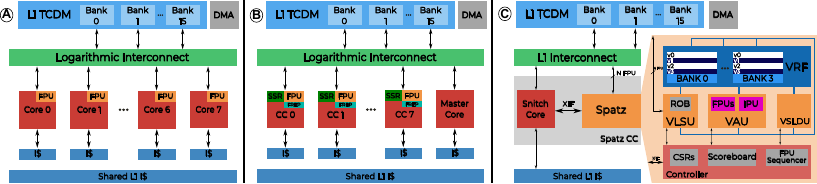}
	\caption{(A) PULP Cluster, (B) Snitch Cluster and (C) Spatz architectures. The first two highlight the multi-core architecture, with Snitch adding the streamers and FREP. Spatz instead showcases the vector unit, placed into a Core Complex. }
	\label{fig:Fig_1}
\end{figure*}

\section{Architectural Background}
This section introduces the architectures that are studied in this paper from a hardware point of view. Two were chosen following the work by Kiamarzi et al.~\cite{10.1145/3801487.3801835}.
\subsection{PULP Cluster}
The Parallel Ultra-Low Power (PULP) Cluster~\cite{7864441} platform is a RISC-V multi-core system. As depicted in Fig.~\ref{fig:Fig_1}A,
the compute cluster is built from eight 32-bit RI5CY cores, each optimized for fast DSP calculations. All cores support ISA extensions targeting energy-efficient digital signal processing, which includes post-incremented Load and Store operations, hardware loops, and packed Single-Instruction-Multiple-Data (SIMD) instructions, enabling efficient acceleration of DSP algorithms. In the configuration used in this paper, each core is paired with its private FPU~\cite{mach2020fpnew}, supporting all 32-bit precision operations, including division/square root.
On the memory side, the hierarchy consists of 128 KiB of single-cycle-latency Tightly-Coupled Data Memory (TCDM), divided into 16 banks and referred to as L1. The latter is a scratchpad memory that requires explicit data management and transfers. Data movement efficiency is ensured by a dedicated Direct Memory Access (DMA) unit.
\subsection{Snitch Cluster}
Fig.~\ref{fig:Fig_1}B depicts the architecture of the Snitch Cluster~\cite{9216552}. The latter is composed of one master core and eight Core Complexes (CC). One complex contains a RV32IMAF, single-issue, single-stage core. This architectural choice has been taken to result in the core being the smallest possible in terms of equivalent gates, using only 9-22kGE~\cite{9216552}. Each core is paired with an FPU~\cite{mach2020fpnew} with a Floating Point Register (FREG) comprised of 32 registers. 

It is possible to configure a custom instruction buffer connected to the FPU with a custom extension called floating-point repetition instruction (FREP). The aforementioned buffer serves the purpose of feeding the FPU with a set of instructions, which will be repeated as set by the programmer. This operation renders the integer core and the FPU subsystem independent, allowing them to work on different instructions without having to synchronize. The integer core and FPU are equipped with three SSRs~\cite{9068465} each. The SSRs are used to configure some streamers that automatically write/read data from the TCDM to a pre-assigned register selected among \texttt{ft0/ft1/ft2} for floating point operations or \texttt{t0/t1/t2} for integer operations. This elides all explicit loads and stores, allowing the instructions to complete with a delay of only the computation cycles. The memory subsystem is configured to have 128 KiB of data memory in the form of 16 banks of TCDM, connected via a wide AXI-bar (512 bits) used by the DMA and a narrow one (32 bits) used by the cores.
\begin{figure*}[t]
	\centering
	\includegraphics[width=1.\textwidth]{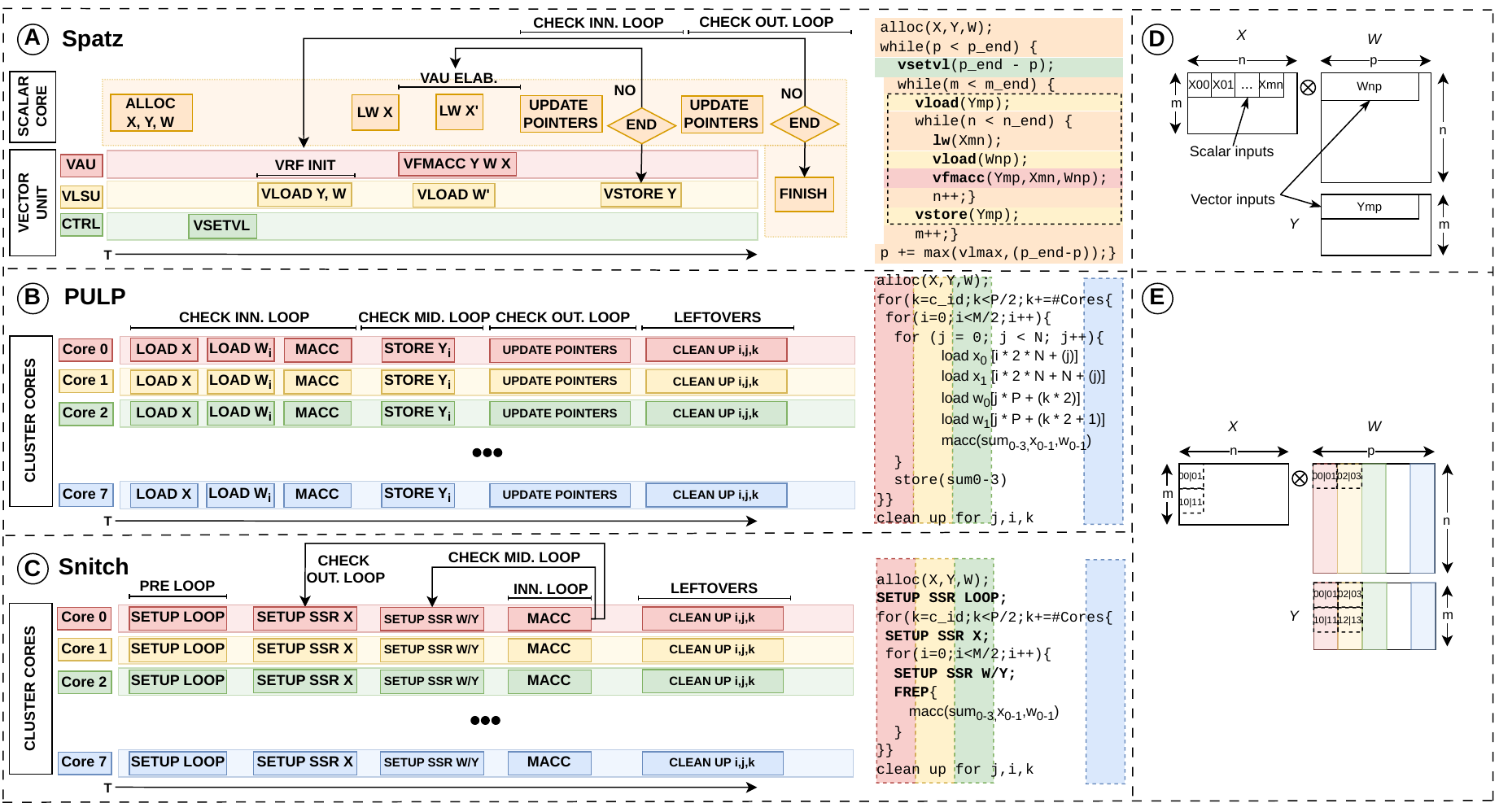}
	\caption{GEMM execution on all three clusters. A) On Spatz, the figure highlights the vector components, while also showcasing the integer core checks. On B) PULP and C) Snitch instead the focus is on the division of the workload on the eight cores, and the various phases of the setup and computation. D-E) Data tile organization on the vector unit and PULP-Snitch Cluster. Time is not to scale.}
	\label{fig:Fig_2}
\end{figure*}

\subsection{Spatz}
Spatz, depicted in Fig.~\ref{fig:Fig_1}C, is a lightweight, configurable Vector Processing Unit (VPU) built on the RVV 1.0 specification~\cite{riscv-v-spec}. It is comprised of a Vector Arithmetic Unit (VAU) handling both integer and floating-point operations, with eight FPUs comprising the Vector Functional Units (VFU); a Vector Register File (VRF), which is shared across all its functional units; a Vector Load Store Unit (VLSU) that is used to perform constant-strided memory accesses as well as indexed gather/scatter patterns and a Vector Slide Unit (VSLDU) that is responsible for data reorganization tasks such as vector slides and moves.

 Spatz functions as a co-processor, and it is integrated with a control Snitch core.
 The latter can dynamically change the configuration of Spatz's Control and Status Registers (CSRs), thus allowing the data sizes (vl and vtype, the active vector length and the element-width/register grouping) to adapt for the execution. The control core also decodes the incoming instructions and offloads them to Spatz via the XIF interface, by accessing the Controller unit. 
 
 In this setup, Spatz and Snitch are paired into a Core Complex (CC) deployed in a shared-L1 cluster comprising sixteen 8 KiB SRAM banks, for a total of 128 KiB. This configuration can be seen in the architecture depicted in Fig.~\ref{fig:Fig_1}C. The memory subsystem works in the same way as the Snitch Cluster, where the control Snitch core performs all data transfers via a dedicated DMA unit.

\section{Comparative Analysis of Computational Models}
Fig.~\ref{fig:Fig_2} showcases the execution of general matrix multiplication (GEMM) on all three designs. This algorithm was chosen as it is a regular workload that can easily showcase the working of the three systems.
\subsection{Spatz Execution Model}
Execution on Spatz is depicted in Fig.~\ref{fig:Fig_2}A and D. The scalar core (Snitch) allocates the needed matrices in memory, then during execution it fetches and decodes vector instructions that are then dispatched to the Vector Functional Units (VFUs). At the beginning of each inner-loop iteration, the vector unit initializes the vector length parameter (VSETVL) and fetches the corresponding Y and W tiles into the Vector Register File (VRF). When the kernel dimensions are small, the computation can be carried out efficiently by leveraging software-configurable data-flow strategies, either input-stationary or weight-stationary, to maximize operand reuse. As kernel sizes grow larger, an outer-product formulation becomes more advantageous: a scalar element drawn from the input tensor is multiplied against a vector slice of the weight tensor, allowing the unit to sustain high throughput by keeping scalar and vector operands pre-loaded in registers. At every iteration, partial accumulation results are written back to memory, while the scalar core is responsible for advancing the loop control state. This execution scheme exemplifies the tight interplay between the scalar and vector cores, as further illustrated by the pseudo-code presented alongside this description.
\subsection{PULP Cluster Execution Model}
To efficiently handle data-parallel workloads on the PULP Cluster a Single Program Multiple Data (SPMD) execution model is employed, where each core operates on a distinct partition of the output tensor using an identical kernel, Fig.~\ref{fig:Fig_2}B and E. This approach achieves scalable throughput by exploiting core-level parallelism, while offering flexible work distribution strategies such as tiling along the M, N, or P dimensions of a GEMM operation. At the innermost level, the micro-kernel follows a canonical load-compute-store pattern: operands are fetched iteratively, fused multiply-accumulate operations are performed, and partial sums are accumulated either in registers or in local storage.
The multi-core cluster architecture further extends this flexibility by allowing the degree of parallelism to be tuned through the number of active cores, and by adapting the work partitioning granularity to match the problem size and structure. This makes the cluster well-suited for a broad range of workloads, including small GEMMs, reduction operations, and non-linear activation functions.
\subsection{Snitch Cluster Execution Model}
The execution of data-parallel workloads on the Snitch Cluster, Fig.~\ref{fig:Fig_2}C and E, closely mirrors the approach adopted for the PULP Cluster, with the same tiling strategy applied along the M, N, or P dimensions of the GEMM operation to distribute work across cores. The key distinction lies in the per-core pre-loop phase that precedes the main computation loops: each core configures its streamers to cover all required memory accesses, including reads from the input X and W matrices and writes to the output Y matrix. In the first initialization phase, pre-loop, each core sets up the loop size and the read/write stride. Once this operation is complete, the integer core enters the outer loop body, where the address of the first matrix is fed to the first data streamer. After this second phase, the core enters the inner loop. This stage is divided into two: the first one is another setup phase of the SSRs, where the second and third streamer are fed the address and offset of the column of the second matrix and of the destination matrix. 

Then, the core programs the FREP to let the FPU repeat independently the MACC operation $n$ times. In the meantime, the integer core goes on handling address calculations and control. This decoupling renders the two units effectively independent: while the integer core prepares the next iteration, the FPU is fed operands directly from the SSR streams, bypassing explicit load instructions entirely. The result is a significant reduction in the number of cycles between consecutive MACC operations, with a corresponding increase in sustained throughput. Together, the SSR and FREP mechanisms shift the majority of the memory and control overhead away from the critical compute path, enabling the Snitch core to reach over half of peak arithmetic utilization with little software overhead.

\subsection{Bottlenecks and Mitigations}
 While the PULP Cluster encounters the classical von Neumann bottleneck, where operations are constrained by memory throughput and the overhead of explicit load and store instructions, the other two designs adopt complementary strategies to mitigate this limitation. In the GEMM kernel, a steady-state iteration on the PULP Cluster requires multiple explicit load operations to fetch the next operand fragments alongside multiple MACC operations to update the accumulators. The example shown in Fig.~\ref{fig:Fig_2}B exhibits an iteration structure with four load operations and four MACC operations in the critical inner loop. Even under ideal scheduling, this implies that a substantial fraction of issue slots and shared-memory bandwidth is consumed by operand delivery rather than arithmetic, making it difficult to sustain more than roughly 50\% compute utilization when the loop is limited by load throughput and TCDM arbitration.

Spatz tackles the same bottleneck through vector execution, improving the balance between control and data movement on one side and arithmetic operations on the other. First, the overhead of instruction fetch, decode, and loop control is amortized across multiple MACC operations, lowering front-end pressure, a cost further distributed across eight FPUs operating in parallel. Second, each vector instruction encodes sufficient micro-operations to keep the FPU datapath busy for extended cycles, ensuring high utilization even at lower instructions per cycle (IPC). Additionally, vector loads and VRF reuse reduce the frequency of explicit operand-delivery instructions relative to the amount of arithmetic performed, particularly in regular kernels where tiles of X and W can be streamed and reused across multiple updates of Y. While the memory system remains a fundamental constraint, vector execution increases the number of useful MACC operations performed per unit of instruction and control overhead, and structures memory access into wider, more regular transfers.
    
The Snitch architecture addresses the von Neumann bottleneck at its source. By offloading operand streaming to the SSR hardware, explicit load and store instructions are eliminated from the inner loop entirely, allowing the FPU to be fed data directly every cycle without stalling on memory access. Simultaneously, the FREP buffer decouples instruction issue from FPU execution, enabling a continuous sequence of MACC operations with minimal integer core involvement. Together, these mechanisms shift data movement off the critical compute path, enabling higher arithmetic utilization compared to PULP within a compact scalar core.

\begin{table*}[t]
\centering
\resizebox{\textwidth}{!}{%
\begin{tabular}{@{}ll cc cc cc *{11}{c}@{}}
\toprule
& & \multicolumn{2}{c}{PULP Cluster} & \multicolumn{2}{c}{Snitch Cluster} & \multicolumn{2}{c}{Spatz} & \multicolumn{11}{c}{Code Patterns} \\
\cmidrule(lr){3-4}\cmidrule(lr){5-6}\cmidrule(lr){7-8}\cmidrule(lr){9-19}
Domain & Workload & Tot. cycles [\#] & FPU Util. [\%] & Tot. cycles [\#] & FPU Util. [\%] & Tot. cycles [\#] & FPU Util. [\%] &
\rotatebox{90}{CI} & \rotatebox{90}{HP} & \rotatebox{90}{LP} & \rotatebox{90}{NPI} & \rotatebox{90}{MI} & \rotatebox{90}{PRAP} & \rotatebox{90}{IMA} & \rotatebox{90}{CD} & \rotatebox{90}{DCF} & \rotatebox{90}{DNL} & \rotatebox{90}{RED} \\
\midrule
\multirow{7}{*}{LA}
 & CHOLESKY Decomp. & \textbf{35989}  & \textbf{16.66} & 36984  & 16.21 & 160595  & 3.73  &            &            & \checkmark &            &            & \checkmark & \checkmark & \checkmark &            &            &            \\
 & LU Decomp.       & \textbf{52525}  & \textbf{7.36}  & 52791  & 7.33  & 185743  & 2.08  &            &            & \checkmark &            &            & \checkmark & \checkmark & \checkmark &            &            &            \\
 & LU Solve         & 15584           & 3.49           & 21026  & 2.59  & \textbf{14847}  & \textbf{3.66}  &            &            &            &            &            &            &            & \checkmark &            & \checkmark & \checkmark \\
 & GEMV             & 6542            & 31.79          & \textbf{2604} & \textbf{79.88} & 6487  & 32.06 & \checkmark & \checkmark &            &            &  & \checkmark &            &            &            &            & \checkmark \\
 & GEMVT            & 7140            & 29.13          & 3386   & 61.43 & \textbf{2620}  & \textbf{79.39} & \checkmark & \checkmark &            &            &  &            &            &            &            &            & \checkmark \\
 & GEMM             & 72412           & 45.25          & 56014  & 58.50 & \textbf{36949} & \textbf{88.68} & \checkmark & \checkmark &            &            &            & \checkmark &            &            &            &            &            \\
 & SVD              & \textbf{507413} & \textbf{31.74} & 975813 & 16.51 & 2192234 & 7.35 &            &            &            &            &            &            & \checkmark & \checkmark &            &            & \checkmark \\
\midrule
\multirow{3}{*}{DSP}
 & FIR & 34025 & 22.60 & 23591 & 32.59 & \textbf{8285}  & \textbf{92.79} & \checkmark & \checkmark &            &            &            & \checkmark &            &            &            &            & \checkmark \\
 & FFT & 6956  & 33.12 & 4306  & 53.51 & \textbf{2644}  & \textbf{87.14} & \checkmark &            &            &            &            &            & \checkmark &            &            & \checkmark &            \\
 & DWT & 35451 & 27.60 & \textbf{27532} & \textbf{35.54} & 35591 & 27.49 & \checkmark &            &            &            &            & \checkmark &            &            &            & \checkmark & \checkmark \\
\midrule
\multirow{5}{*}{ML}
 & K-MEANS        & 114216 & 12.59 & \textbf{71313} & \textbf{20.16} & 403393 & 3.56 &            &            & \checkmark & \checkmark &            & \checkmark &            &            & \checkmark &            &            \\
 & CONV3x3         & 6302   & 16.07 & 2284   & 44.33 & \textbf{2189} & \textbf{46.25} &            & \checkmark &            &            &            & \checkmark &            & \checkmark &            &            & \checkmark \\
 & SVM EXP BILL  & 70250  & 20.48 & 71066  & 20.25 & \textbf{24573} & \textbf{58.55} & \checkmark &            &            &            &            & \checkmark &            & \checkmark &            &            & \checkmark \\
 & SVM EXP CANC  & 381814 & 33.49 & 258939 & 49.39 & \textbf{144120} & \textbf{88.74} & \checkmark &  &            &            &  & \checkmark &            & \checkmark &            &            & \checkmark \\
 & RANDOM FOREST & \textbf{13608} & 0      & 25899  & 0     & 80398  & 0 &            &            &            &            &            &            &            & \checkmark &            &            &            \\
\midrule
\multirow{7}{*}{LLM}
 & ENCODER   & 209152    & 11.75          & \textbf{61056} & \textbf{40.25} & 106752    & 23.02 & \checkmark &            &            &            & \checkmark & \checkmark &            &            &            &            &            \\
 & GELU      & 3038592   & 48.53          & 15754944 & 9.36 & \textbf{2788800}   & \textbf{52.87} & \checkmark & \checkmark &            & \checkmark &            &            &            &            &            &            &            \\
 & RESIDUAL  & 177984    & 13.81          & \textbf{46752} & \textbf{52.57} & 92448    & 26.58 &            &            &            &            & \checkmark & \checkmark &            &            &            &            &            \\
 & SOFTMAX   & 58724352 & 16.43 & 136335360 & 7.08 & \textbf{36970496} & \textbf{26.10}  &            & \checkmark &            & \checkmark &            &            &            &            &            &            & \checkmark \\
 & MATMUL    & 121074048 & 46.77  & 90170496 & 62.80 & \textbf{59897664}  & \textbf{94.53} & \checkmark & \checkmark &            &            &            & \checkmark &            &            &            &            &            \\
 & ATTENTION & 19309920 & 8.91 & \textbf{16150416} & \textbf{10.66} & 28675968  & 6.00  &            &            &            &            &            &            &            &            &            & \checkmark &            \\
 & LAYERNORM & 933888   & 21.07 &  \textbf{434264} & \textbf{45.30}  & 519424   & 37.88  &            &            &            &            &            & \checkmark &            &            &            &            & \checkmark \\
\bottomrule
\end{tabular}%
}
\caption{Comparison of PULP Cluster, Snitch Cluster, and Spatz across LA, DSP, ML, and LLM kernels. Bold values indicate the best result per row (lowest cycle count / highest FPU utilization). Code Patterns: CI = Compute-Intensive; HP = High Parallelization; LP = Low Parallelization; NPI = Non-Pipelined Instruction (e.g., div/sqrt); MI = Memory-Intensive; PRAP = Partial Regular Access pattern (e.g., codes with section with linear memory access); IMA = Irregular Memory Access; CD = Control-Dominant; DCF = Divergent Control Flow; DNL = Dependent Nested Loop; RED = Reduction.}
\label{tab:Table_2}
\end{table*}

\section{Application Performance Analysis}
Table~\ref{tab:Table_2} reports the extracted cycles and FPU Utilization for the three systems in twenty-two applications. This section will go over all of them, explaining the different performance of the three platforms.
\subsection{Linear Algebra}
Linear Algebra kernels contain an ideal mix of computing and control sections.
It can be clearly seen that for workloads with a high computing usage, Spatz outperforms Snitch and PULP Cluster.
Instead, in the control-dependent ones PULP Cluster outperforms the other two, as its integer cores are more powerful compared to the Snitch ones.
\subsubsection{Cholesky Decomposition} We evaluate this kernel on a $64\times64$-element symmetric positive-definite matrix. On the PULP Cluster, both diagonal computation and column updates are distributed across cores, relying on local accumulation followed by a final reduction; however, the outer loop's frequent synchronizations cap the FPU utilization at 16.66\%. Snitch Cluster implements the same code as the PULP Cluster, restricting the usage of the SSRs to Partial Regular Access Sections such as the diagonal factorization and column/panel computation. Nevertheless, the architecture reaches a slightly lower utilization compared to PULP, 16.21\%. On Spatz, the diagonal and update steps are instead vectorized through fused multiply–accumulate operations and vector reductions, yet the outer loop's inherently sequential nature keeps utilization low, at just 3.73\%.
\subsubsection{LU Decomposition} We target a $64\times64$ matrix using pivot selection combined with Gaussian elimination. On PULP, only the elimination phase is parallelized across cores through row updates, while pivot search remains inherently serial; this phase synchronization restricts utilization to 7.36\%. The Snitch implementation is identical, where SSRs are only used during elimination. This improves the performance and closes the gap to PULP, allowing the design to reach a 7.33\% utilization. On Spatz, both phases are vectorized, but loop-carried dependencies still constrain utilization to a mere 2.08\%.
\subsubsection{LU Solve} We solve $Ax = b$ through forward and backward substitution for a $64\times64$ A matrix. On PULP, both phases are parallelized using local dot-product accumulation followed by a final reduction, yet synchronization overhead together with loop-carried dependencies caps utilization at 3.49\%. On Snitch the kernel is implemented without the usage of any specific acceleration, showcasing the limitations of the scalar core. The performance is bottlenecked by the control and reduction, tanking its performance to a mere 2.59\% utilization. On Spatz, the computation instead relies on fused multiply–accumulate operations and vector reductions that are carried out by a parallel reduction tree, letting the design achieve 3.66\% utilization and perform slightly better than the other two.
\subsubsection{GEMV/GEMVT}For both GEMV and GEMVT we perform the computation on a $128\times128$ matrix. On PULP, parallelizing over rows produces comparable behavior for the two kernels, with utilization of 31.79\% for GEMV and 29.13\% for GEMVT. On the  Snitch Cluster, the workload is split between the cores by rows like in the previous implementation, resulting in memory accesses on different banks for each core and almost no contention. This pattern, in addition to the use of the streamers that effectively elide all memory operations, results in a 79.88\% utilization on GEMV and 61.43\% on GEMVT. On Spatz, GEMV depends on strided vector loads and the final reduction phase caps the maximum utilization, reaching 32.06\%. Instead, on GEMVT it obtains 79.39\%, as there is no final reduction.
\subsubsection{SVD} We profile SVD on a $64\times32$ matrix by computing $G = A^TA$, applying Jacobi eigendecomposition to obtain V, and finally recovering $U = AV\Sigma^{-1}$. On PULP, both the G computation and the final stage parallelize well, whereas Jacobi eigendecomposition requires a global reduction at each round, bringing overall utilization to 31.74\%. This kernel represents another of Snitch's streamers limitations, thus obtaining a 16.51\% utilization, as the synchronization and reduction represent a clear bottleneck. On Spatz, all three stages are vectorized using strided loads and stores, yet the associated control and synchronization overhead reduces utilization to 7.35\%.
\subsection{Digital Signal Processing}
DSP kernels comprise a realistic picture for applications in embedded systems. Here we can clearly see that Spatz has a clear advantage in two out of three profiled workloads.
\subsubsection{FIR} In the Finite Impulse Response (FIR) filter we evaluate a 64-tap filter with an input data comprised of N=1024 elements, ignoring the boundaries and producing a 961-element output. On the PULP Cluster, scalar operand delivery dominates the inner loop, requiring four loads plus two MACC operations per iteration. Because the accumulation creates a loop-carried dependency, a one-cycle stall is introduced, capping FPU utilization at 22.60\%. 

Instead, on Snitch, the usage of SSRs and the linear memory access pattern relaxes the cycle latency, resulting in lower total cycles and a significantly higher FPU utilization (32.59\%). Spatz instead adopts a sliding-window vector approach: each iteration brings a signal chunk into the VRF, shifts the window through slide operations, and executes the MACC. Since load, slide, and compute are overlapped across the VLSU, VSLDU, and VAU, the load latency is effectively hidden, allowing utilization to remain close to the peak at 92.79\%.
\subsubsection{FFT}
We evaluate a 512-point complex FFT, a workload that is compute-intensive yet suffers from irregular memory access, driven by staged butterfly operations and the associated data rearrangements. Its dominant step consists of about 10 memory operations against 8 floating-point operations. Under this profile, PULP attains 33.12\% utilization; Snitch a 53.51\% utilization, improving from PULP but still suffering from the frequent re-synchronization; Spatz pushes the performance to 87.14\% by leveraging vector chaining together with concurrent execution across its vector units.
\subsubsection{DWT} This workload employs a 40-coefficient filter across an input with 1000 elements, producing 1155 outputs over four decomposition levels. Unlike the standalone FIR kernel, here we perform boundary handling, which reduces compute intensity.

Nonetheless, since DWT is built from two FIR filters computed in parallel, our implementation is able to offset the accumulation latency that limits the standalone FIR kernel, yielding higher average utilization overall: 27.60\% on PULP and 35.54\% on Snitch. Snitch can map the two independent accumulation chains directly onto its SSR streams, allowing the FREP loop to keep the FPU saturated despite the reduced compute intensity. The Spatz implementation, by contrast, is burdened by a chain of slides with dependencies on one another and by the use of large strided loads, obtaining a low 27.49\% utilization, comparable to PULP but well below Snitch's 35.54\%.
\subsection{Machine Learning}
The increasing demand for Machine Learning on embedded systems has augmented the need for efficient computation~\cite{11314213}. These kernels have increased their importance over the years, and thus are needed for a clear picture of all workloads.
\subsubsection{K-means} The dataset is comprised of 8 clusters and 100 objects characterized by 8 features. For PULP, the nearest-center search parallelizes effectively, yet cluster updates depend on reductions, shared-state updates, and long-latency divisions, capping utilization at 12.59\%. The same implementation is run on Snitch Cluster, with the only difference of the usage of SSRs and the FREP loop for the euclidean distance calculation. This helps reduce cycles during heavy computation phases and reach a utilization of 20.16\%. On Spatz, distance computation vectorizes efficiently, whereas membership updates and cluster statistics demand scalar control along with per-element decisions, bringing utilization down to 3.56\%.
\subsubsection{Convolution} This benchmark computes a 3×3 convolution over a $32\times32$ matrix input, using unit stride and no padding. On PULP, parallelization is carried out over output rows, but the inner loop remains load-intensive and carries an accumulation dependence chain, resulting in 16.07\% utilization. Instead, the kernel for the Snitch Cluster is fully optimized to use SSRs in combination with the FREP instruction buffer, with minimal cycle latency between compute instructions and no synchronization between loop iterations, resulting in a 44.33\% utilization, near the maximum obtained by Spatz. The latter uses a sliding-window vector implementation that overlaps loads, slides, and MACC operations while sidestepping reductions altogether, reaching 46.25\% utilization, higher than both other designs.
\subsubsection{Support Vector Machines} We profile an RBF-kernel SVM employing Schraudolph's exponential approximation~\cite{10.1162/08997669930001646} on the BILL~\cite{uci_banknote2026} and CANC~\cite{uci_breastcancer2026} datasets. The model relies on 100 support vectors and evaluates 100 decision points, with feature dimensions of 4 for BILL and 30 for CANC. 

On PULP, parallelization is carried out across support vectors, delivering 20.48\% utilization for BILL and 33.49\% for CANC. Meanwhile on Snitch, the usage of SSRs is restricted to only the CANC dataset. On the BILL dataset, the overhead is too high to favorably utilize them. This results in a 20.25\% utilization for BILL. Instead, with the usage of streamers and the CANC dataset, it reaches a 49.39\% utilization. This outcome highlights the relevance of the streamers, as with the BILL dataset the two designs perform identically, while with CANC data Snitch achieves a 47\% higher utilization compared to PULP. 

On Spatz, strided vector loads together with vector MACC operations speed up both distance evaluation and the exponential approximation, though utilization fluctuates between datasets depending on feature dimensionality and control-flow overhead, therefore resulting in 58.55\% and 88.74\% utilization.

\subsubsection{Random Forest} This benchmark is governed primarily by integer comparisons and conditional branches. Thus, it helps us compare the two scalar cores (RI5CY and Snitch). We consider an ensemble of six decision trees, each containing between 280 and 300 nodes. On PULP and Snitch, each tree is allocated to a distinct core, while on Spatz the integer control flow runs entirely on the Snitch core, leading to zero FPU utilization on all platforms. Here it can be clearly seen how the Snitch integer core underperforms compared to the RI5CY core. Furthermore, the code highlights how scaling from one Snitch Core to a cluster of eight, with six cores utilized to compute, does not reduce the cycles by the ideal factor six, but only by a factor equal to 3.10.
\subsection{Large Language Models} The GPT-2 kernels from the reference implementation~\cite{karpathy_llmc} use a dataset with batch size 4 and sequence length 64. We target GPT-2 124M with $C = 768$, 12 attention heads,
and a vocabulary of 50,257 (padded to 50,304), yielding a QKV projection output of 2304 ($3 \times C$). The channel dimension $C$ is the main parallel axis. Kernel-specific control flow, reductions,
and long-latency operations limit peak utilization of the compute elements.
\subsubsection{Encoder/Residual}These two benchmarks are dominated by regular memory access and low control dependence. Therefore, they are memory-bound. On PULP, the overall performance is bottlenecked by the memory latency, yielding an 11.75\% and 13.81\% utilization. Snitch, instead, thrives in these kinds of applications, as the memory accesses are linear and optimized by the streamers, obtaining a 40.25\% and 52.57\% utilization. The results on Spatz outline an improvement compared to PULP, thanks to the mitigation obtained by overlapping the computation with VLSU activity, resulting in 23.02\% and 26.58\% utilization, bottlenecked by the VLSU maximum bandwidth.
\subsubsection{GELU} This kernel is dominated by division from the $\tanh$ and activation calculations. Here, all designs are constrained by the division computation, resulting in a 48.53\% (PULP), 9.36\% (Snitch) and 52.87\% (Spatz) utilization. The Snitch Cluster is burdened as it is not possible to optimize the compute-intensive and synchronization steps, thus losing a substantial part of performance compared to the other two designs.
\subsubsection{Softmax} Softmax's bottlenecks are the divisions and reductions, similarly to GELU. Here, we clearly see that PULP Cluster is only limited by the divisions and can reach 16.43\% utilization. Meanwhile, Spatz improves the overall utilization by using its parallel reduction tree, and thus the reduction instruction, obtaining a 26.10\% utilization. Finally, Snitch is burdened by reductions, divisions and  the absence of SSRs usage thus obtaining a mere 7.08\% utilization.
\subsubsection{Matrix Multiplication} The Matmul kernel is compute-intensive and follows the behavior of the GEMM kernel explained in Section 4. The results confirm this behavior, with PULP obtaining a 46.77\% utilization, Snitch 62.80\% and Spatz 94.53\%.
\subsubsection{Attention} This benchmark exhibits irregular, lower-triangular computation due to causal masking,
leading to time-dependent loop bounds that reduce vectorization and parallel scheduling efficiency. The result is that Snitch Cluster outperforms the others, with a 10.66\% utilization. PULP Cluster implements the same code and obtains a slightly lower utilization, 8.91\%. Spatz instead is bottlenecked by the irregular computation and obtains a 6.00\% utilization.
\subsubsection{Layernorm} Layernorm is dominated by reductions, in a similar manner to Softmax. This is clearly seen with the PULP Cluster result, which obtains a slightly higher utilization compared to the latter kernel (21.07\%). The Snitch implementation instead can be fully optimized by the usage of SSRs and two FREP calls. The latter reduces almost to zero the control overhead, removing the integer control bottleneck from the FPU operations. This allows the design to reach a 45.30\% utilization. Spatz uses a parallel reduction tree, helping the vector accelerator reach a 37.88\% utilization.

\subsection{Programming Effort and Complexity}
Table~\ref{tab:Table_3} reports the programming effort of the three architectures through three quantitative metrics (the total lines of code for all profiled kernels, considering only the sections of the actual workload; the number of lines written in assembly (ASM); and the number of synchronization primitives used, e.g., barriers or FPU fences) and four qualitative aspects, denoting with a check mark the techniques that the programmer must apply manually. The PULP Cluster is programmed directly in C, hence only 0.71\% of its lines are ASM, mostly minor optimization on some specific sections, while its code length and primitive count lie between those of Snitch and Spatz. Loop unrolling and barriers increase the coding complexity, although not substantially.

\begin{table}[t]
\centering
\begin{tabular}{@{}lccc@{}}
\toprule
\textbf{Aspect} & \textbf{PULP} & \textbf{Snitch} & \textbf{Spatz} \\
\midrule
Total lines of code          & 1837 & 1366  & 2892  \\
ASM lines (\%)               & 0.71 & 14.28 & 37.28 \\
Synchronization primitives   & 48   & 58    & 27    \\
\midrule
Loop unrolling               & \checkmark & $\times$   & \checkmark   \\
Manual pipelining            & $\times$   & $\times$   & \checkmark \\
Manual address patterns      & $\times$   & \checkmark & $\times$   \\
Control/compute sync.        & $\times$   & \checkmark & \checkmark \\
\bottomrule
\end{tabular}
\caption{Comparison of programming effort across the three cluster architectures. A check mark denotes a technique that must be applied manually by the programmer.}
\label{tab:Table_3}
\end{table}

Snitch and Spatz are programmed in a mix of C and RISC-V assembly. On Snitch, ASM lines rise sharply to 14.28\%. Besides inter-core barriers, FPU barriers are needed to synchronize the control (scalar core) and compute (FPU) sides after a decoupled computation section, increasing the overall number of primitives and debugging complexity. Furthermore, the need to manually compute the address patterns considerably increases the complexity and the required knowledge of the specific workload. The lower line count, 26\% lower than PULP, is due to Snitch kernels not requiring explicit loop unrolling: on PULP, computing sections are unrolled (e.g., 2 loads, 2 operations, 2 stores), whereas on Snitch memory operations are elided from the instruction stream, so only the compute instructions contribute to the line count.

Spatz reaches the maximum in both code lines and ASM percentage, with 2892 lines and 37.28\%. This increase is due to the need unroll the loops and manually pipeline the VAU, VLSU, and VSLDU, which requires kernel-specific tuning. The primitive count instead drops, since only control/compute synchronization between the scalar core and the vector unit is needed.

Although Snitch has the lowest line count, its ASM share, manual address-pattern computation, and synchronization make it more demanding to program than PULP. Spatz is even more demanding: it has the highest line count and ASM share, and it requires manual pipelining. Overall, a higher degree of micro-architectural specialization demands more software effort to exploit all the potential performance gains.

\section{Physical Implementation}
In this section we are going over the physical implementation of the three designs, from the tools and methodologies used to the results. The frequency, power and area results are then used to depict a fair PPA comparison on all the previously profiled workloads, giving a complete picture of the three architectures.
\subsection{Experimental Setup}         

\begin{figure*}[t]
	\centering
	\includegraphics[width=1\textwidth]{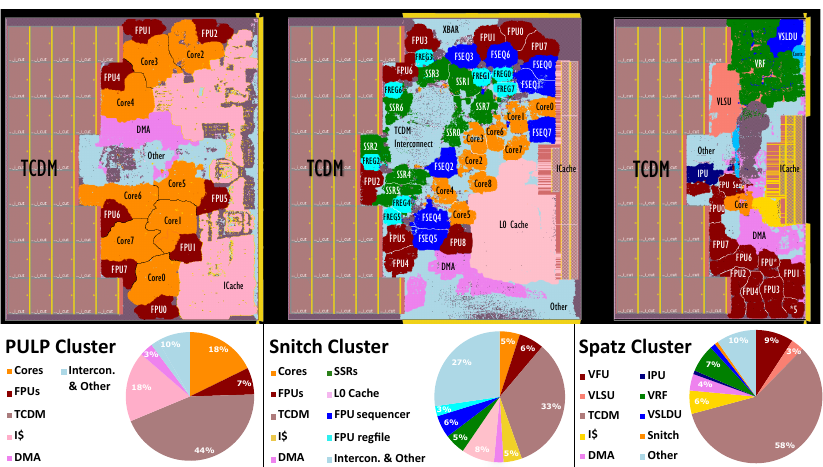}
	\caption{Post-Layout area breakdown. The PULP Cluster layout is the least complicated, with only the Cores, FPU, Memory, DMA and Interconnect highlighted. The Snitch Cluster adds some complexity and overall area with the SSRs, more complex FPSS and L0 Cache. Then Spatz, the smallest design, highlights all the components implemented in a vector architecture.}
	\label{fig:Fig_3}
\end{figure*}
The physical implementation for the three designs has been conducted in GlobalFoundries GF12 technology. This node is a prime example of cutting-edge technology.

The synthesis flow has been implemented with Synopsys Design Compiler NXT v2025.06 in topographical mode, using the SSPG 0.72V -40°C as the synthesis corner. The Place and Route has been performed using Cadence Innovus v23.33, considering all worst-case corners.

The floorplan for the P\&R has been kept consistent between all three designs, setting the height to $800$ $\mu$m and adapting the width to target a 70\% total utilization, while targeting a 500 MHz working frequency.
The memory layout spans the full height of the floorplan, as all designs feature 128 KiB of TCDM implemented in 16 banks composed of 32 identical memory cuts.

The Post Layout simulations have been carried out using QuestaSim v2025.3.
Finally, the power estimation has been carried out with Synopsys PrimeTime 2025.06, using the TT 0.80V 25°C corner.

\begin{table}[h!]
\centering
\begin{tabular}{@{}lcc@{}}
\toprule
& Frequency [MHz] & Area [mm\textsuperscript{2}] \\
\midrule
PULP Cluster   & 810 & 0.396 \\
Snitch Cluster & 827 & 0.456 \\
Spatz          & \textbf{846} & \textbf{0.299} \\
\bottomrule
\end{tabular}
\caption{Maximum operating frequency and area, obtained by a topographical synthesis, for PULP Cluster, Snitch Cluster, and Spatz.}
\label{tab:Table_4}
\end{table}

\subsection{Physical Implementation Results}  
To have a fair comparison of the three clusters, the maximum working frequency has been obtained by a topographical synthesis, using the same floorplan as the P\&R. We have also applied a 5\% uncertainty on all clocks.

The closing frequencies and the areas have been reported in Table~\ref{tab:Table_4}. For all, the critical path is from a TCDM bank to a specific memory register. Spatz closes at a frequency 2.3\% higher than Snitch and 4.4\% higher than PULP, as its critical path goes through the L1 Interconnect, that is not logarithmic like the other two and thus it consumes less of the critical timing. Snitch's critical path ends in the SSR registers, passing through the more complex L1 interconnect, thus closing at a lower frequency compared to Spatz. PULP, instead, closes at the slowest frequency, as its critical path goes through the LSU, EX-stage of the core and ends in the ID-stage. 

Furthermore, as for the area at these frequencies, Spatz occupies the least, 34.4\% less than Snitch and 24.5\% less than PULP. The Snitch Cluster is the biggest design, while PULP is a midpoint between the two. 

Figure~\ref{fig:Fig_3} showcases the layout of all the designs after Place \& Route. As presented before, the floorplan is modular and follows the same rules for all three designs. The height is capped by the TCDM at 800 $\mu $m. The width is instead chosen to achieve near 70\% total utilization. This is done to depict how a real chip could be structured using the designs. Each design's layout is also paired with a corresponding area breakdown.

PULP Cluster ($670\mu $m $\times$ $800\mu $m): This design's computing area is dominated by the cores, which represent 18\% of the total area. The difference from the much smaller Snitch cores is due to the more complex design, with four pipeline stages against only one. The FPUs occupy just 7\% of the area, while a significant portion, 62\%, is occupied by the memory subsystem: 44\% by the TCDM and 18\% by the I\$. 

This area partition highlights the flexibility of the PULP Cluster, where the control and memory occupy most of the area, while the computational part represents a much smaller fraction;

Snitch Cluster ($780\mu$m $\times$  $800\mu$m): For the computing units, the FPU subsystem (FPSS) represents most of the computing area with 15\% of the total. Under this category, the FPUs themselves take up 6\%, the Sequencers 6\% and the Register File 3\%. The FPSS is considerably more complex than the one of the other two designs to accommodate the streaming and FPU decoupling capabilities of the specific architecture. 

The nine integer Snitch cores account for just 5\% of the total area. Moving on to the SSRs, which are specific to this design, we can clearly see that they contribute to the total area substantially, by 5\%, almost equal to the cores themselves. Moving on to the memory, the TCDM and I\$ take up 38\% of the total area, while the L0 cache adds a substantial 8\% to the area count. The DMA, the last sub-system, utilizes 2\% of the total. The remaining area, 27\%, is taken up by the most important interconnections (TCDM, AXI and XBAR) by 6\%, and other modules and connections (21\%). 

The area division highlights the more specialized structure of the Snitch Cluster, which thus occupies more area than PULP, with the control cores and the TCDM using a smaller percentage compared to the other designs;

Spatz ($510\mu$m $\times$ $800\mu$m): The area consumption of Spatz is dominated by the memory subsystem (TCDM and I\$), comprising 64\% of the total. Moving on to the computing system, we can clearly see that the VFU, composed of eight FPUs, is the second largest block, using 9\% of the area. The VRF also plays an important part, with 7\% of the total area. Other units, such as the VLSU, VSLDU, Controller and IPU take up under 4\% of the area each. The DMA subsystem, part of the Cluster, takes up 4\% of the total area. As mentioned before, the Snitch integer core is designed to occupy as small an area as possible, 0.81\%. 

This small fraction emphasizes the highly specialized structure of Spatz, where most of the computational area is taken up by the VFU, and the other critical component is the VRF.

\begin{figure*}[t]
	\centering
	\includegraphics[width=1\textwidth]{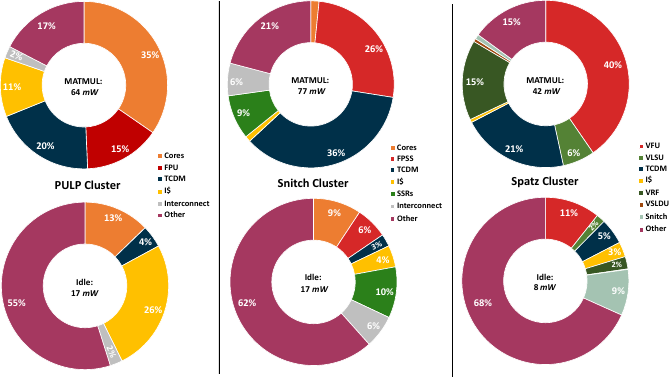}
	\caption{Post-Layout Power breakdown of all the Clusters, considering both a load and idle case. The Figure clearly highlights how, with the same frequency, Spatz consumes almost half of the power of the other two in both cases.}
	\label{fig:Fig_4}
\end{figure*}
\subsection{Power Analysis}
To ensure a fair comparison, all designs are taken through P\&R targeting the same 500 MHz operating frequency, which we selected as a balanced point between performance and power consumption.
We then report a coarse-grained power breakdown for all
platforms under idle and max load conditions in Figure~\ref{fig:Fig_4}. 

The idle condition is obtained by letting one core iterate in an empty for loop, while other cores and units idle. Meanwhile, the max load condition represents a workload where all three clusters obtain a high utilization, relative to their maximum obtained in the twenty-two kernels. The latter is the Matrix Multiplication. With the two distinct power results, we performed a linear interpolation/extrapolation on the FPU utilization to obtain a power estimation for all the workloads presented in Table~\ref{tab:Table_2}.

As for the PULP Cluster, the idle case consumes 17\,mW, with most of the power budget taken from the I\$. In the meantime, the matrix multiplication workload leads to a total power of 64\,mW, with the most significant contributions from the scalar cores and the TCDM.  

As for the Snitch Cluster, the idle case still consumes 17\,mW similarly to PULP, but with a different partition between elements. Instead, in the load case, we see that the FPUs (26\%), TCDM (36\%), SSRs (9\%) and Interconnect (6\%) take up most of the 77\,mW budget, while the cores give a negligible contribution. It can be inferred that the more specialized architecture, with the addition of the streamers, results in a higher required power budget.

In the idle scenario Spatz consumes only 8.36\,mW, thanks to its far less complex design. Under max load Spatz reaches a total power of 42\,mW, with the majority of consumption concentrated in the vector unit, particularly in the vector register file (15\%) and vector functional units (40\%). A significant part is also taken up by the VLSU (6\%) and the TCDM (21\%). 

This behavior is typical of specialized architectures, where most of the power budget is devoted to highly optimized computational units. Instead, in the other two architectures the fraction of power consumed by the compute units is smaller, with the control cores and TCDM taking a higher cut, reflecting the associated overheads required to support a wider range of workloads and a streaming architecture.
\begin{figure*}[t]
	\centering
	\includegraphics[width=1\textwidth]{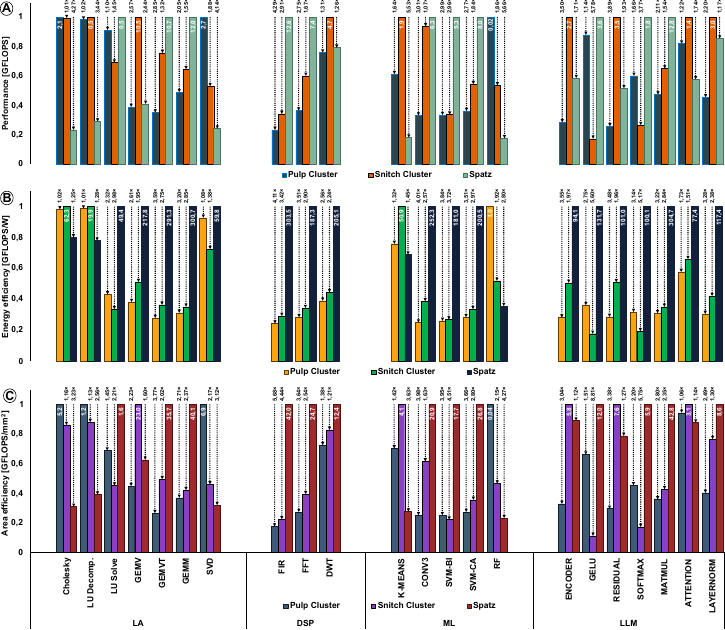}
	\caption{Post-layout PPA comparison between the three architectures. (A) Performance with the maximum frequency (GFLOPS); (B) Energy Efficiency at 500MHz (GFLOPS/W); (C) Area Efficiency (GFLOPS/mm$^2$). All bars are normalized to the maximum observed value. *For Random Forest (RF), values are reported in GOPS.}
	\label{fig:Fig_5}
\end{figure*}

Nonetheless, at the same frequency Spatz consumes almost half of the power compared to the other two, resulting in an overall higher energy efficiency.  PULP and Snitch Cluster instead are tied in the idle case, while for the load case Snitch uses 20.3\% more power compared to PULP. This highlights how the SSRs, more complex Interconnect and FPU subsystem, due to its streaming capability and need of a more complex sequencer structure, contribute significantly to the overall power.

\section{PPA Analysis}
Figure~\ref{fig:Fig_5} compares the three platforms in terms of performance (A), energy efficiency (B), and area efficiency (C), expressed in GFLOPS, GFLOPS/W, and GFLOPS/mm$^2$, respectively. The frequency and area have been taken from the maximum value extracted from the synthesis, while the power from the post P\&R simulations as explained in Section 6.3. To highlight the relative differences across applications and platforms, all measurements are normalized to the highest value recorded in the corresponding category.

The physical implementation results reshape this functional picture, but very unevenly. Maximum frequency has almost no effect: Spatz closes timing only 2.3\% and 4.4\% above Snitch and PULP, so the performance ranking in Fig.~\ref{fig:Fig_5}A essentially follows the cycle counts of Table~\ref{tab:Table_2}, and changes only where two designs were already within about 1\% in cycles (e.g., LU Decomposition). Area and power, instead, change the picture substantially. Spatz is 24.5\% smaller than PULP and 34.4\% smaller than Snitch, and at 500\,MHz it consumes 51\% less power than both in idle (8.36\,mW against 17\,mW), and 34\% and 45\% less than PULP and Snitch, respectively, at max load. As a consequence, being faster in cycles does not imply being more efficient. On Encoder, Residual, Attention, and Layernorm, Snitch needs 1.2-2.0$\times$ fewer cycles than Spatz, yet Spatz is 1.5-2.4$\times$ more energy-efficient than Snitch. Conversely, on Convolution 3$\times$3 and GELU, Spatz needs only 4-9\% fewer cycles than the best competing design, but is 2.5-2.6$\times$ more energy-efficient. The area reduction plays a similar role: on Layernorm, Snitch is 1.17$\times$ faster, but Spatz is 1.3$\times$ more area-efficient. These advantages are not unconditional, though: when Spatz utilization collapses, as on K-means (3.56\%), its lower power no longer compensates for the loss in performance, and both Snitch and PULP become more energy-efficient.
\begin{figure*}[t]
	\centering
	\includegraphics[width=1\textwidth]{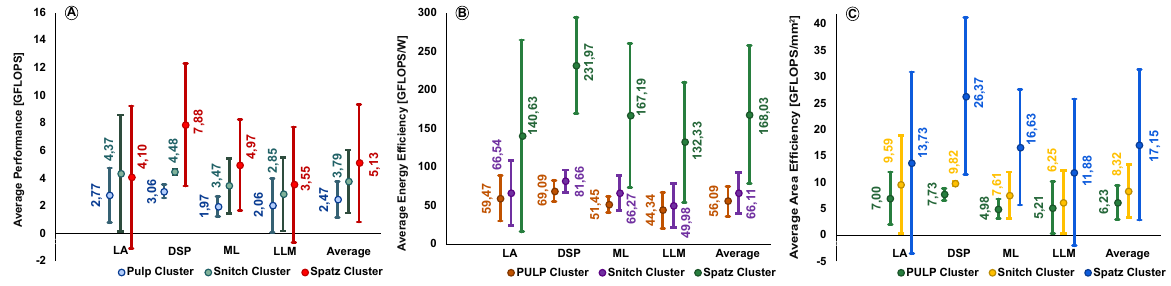}
	\caption{Category-level aggregation of post-layout PPA results for the three architectures. (A) Average performance (GFLOPS), (B) average energy efficiency (GFLOPS/W), and (C) average area efficiency (GFLOPS/mm$^2$) per application category, spanning all applications. Error bars denote mean $\pm$ standard deviation.}
	\label{fig:Fig_6}
\end{figure*}

This can be explained by the different relationship between computation and power consumption in the two architectures. In multi-core processors, the von Neumann bottleneck can become a significant contributor to power consumption when the available parallelism is insufficient to translate additional processing resources into proportional performance improvements. Streamer processors mitigate this effect, but vector processors provide a more direct relationship between useful computation and energy consumption, allowing their power usage to scale more closely with the amount of work effectively performed.

Figure~\ref{fig:Fig_6} highlights the aggregated results for each application domain. From the performance graph, Fig.~\ref{fig:Fig_6}A, it can be clearly seen that in compute-intensive domains such as DSP and ML, Spatz has an average performance higher than the other two designs, and it also reaches the maximum GFLOPS in all domains. Instead, for LA, the Snitch Cluster outperforms the other two on average. PULP Cluster ranks last in both average and maximum GFLOPS. Overall, Spatz achieves on average $2.07\times$ the performance of PULP and $1.35\times$ that of Snitch.

These averages, however, hide the most distinctive trait of Spatz: its very large variability. The error bars in Fig.~\ref{fig:Fig_6}A show that Spatz has the widest spread in performance, and in almost every domain its worst kernel is the worst of the three designs, falling even below PULP. The same trend is visible in Table~\ref{tab:Table_2}: excluding Random Forest, Spatz FPU utilization ranges from 2.08\% (LU decomposition) to 94.53\% (Matmul), a spread of about $45\times$, against about $31\times$ for Snitch and $14\times$ for PULP. Vector execution thus delivers the best results only when the kernel matches its regular, data-parallel structure, while sequential or control-dominated kernels leave the vector unit largely idle.

Energy efficiency shifts the overall picture. Thanks to its lowest power consumption, Spatz is the best-performing design on average and for the maximum value, and its minimum point is also higher than the average of the other two designs in every domain except LA, where the high risk does not pay off. On average, Spatz is $2.99\times$ more efficient than PULP and $2.54\times$ more efficient than Snitch.

As for the area efficiency, Fig.~\ref{fig:Fig_6}C, the picture again favors Spatz, thanks to its smaller footprint: as reported in Section 6.2, it uses 34\% less area than Snitch and 24\% less than PULP. Spatz performs better on average and for the maximum in all domains, being $2.75\times$ more efficient than PULP and $2.06\times$ more efficient than Snitch on average. Nonetheless, the variability remains: on kernels such as the LA factorizations, Spatz can fall well below the other designs even in area efficiency.

\section{Workload Specific Guidelines}
From Figs.~\ref{fig:Fig_5}-~\ref{fig:Fig_6}, we extract guidelines for specific classes of workloads, referring to the three architectures by their execution paradigm: vector (Spatz), streaming (Snitch Cluster), and multi-core (PULP Cluster).

Compute-intensive (CI) and high-parallelization (HP) kernels, with high arithmetic intensity and regular memory accesses, favor vector execution, as control and instruction overheads are amortized across multiple data elements and functional units.

Memory-intensive (MI) and partial-regular-access-pattern (PRAP) kernels are instead constrained by operand delivery and memory-system behavior, and thus streaming performs best. Vector execution partially alleviates these limits through wider and more regular memory accesses, whereas multi-core execution pays the full cost of operand delivery and is competitive only in a handful of cases.
Irregular Memory Access kernels (IMA) are well suited for multi-core execution on the PULP Cluster.

Reduction (RED) and dependent nested-loop (DNL) kernels introduce loop-carried dependencies that restrict parallelism, increase streaming overheads, and reduce the effective vector length. No paradigm is a clear winner, so the choice requires ad hoc reasoning.

Control-dominant (CD) and divergent-control-flow (DCF) kernels disrupt lockstep execution and reduce control amortization, limiting the benefits of vector and streaming execution. Multi-cores systems reveal themselves as a good fit for these kinds of kernels, although in some applications vector and streaming execution can close the performance, energy and area efficiency gap.

Finally, for low-parallelization (LP) kernels, both multi-core and streaming achieve better performance, although mapping constraints and micro-architectural bottlenecks call for case-by-case analysis.

\section{Conclusions}
In this paper, we compared multicore, streaming, and vector execution paradigms through functional and architectural evaluation, post-layout power, performance, and area analysis, and physical implementation of the PULP Cluster, Snitch Cluster, and Spatz in GlobalFoundries 12 nm technology. Our contributions include a systematic comparison of the three architectures, a mapping between kernel characteristics and execution paradigms, and a quantitative assessment of their programming effort. The results show that no single paradigm performs best across all workloads. Spatz has the smallest area footprint and achieves the highest operating frequency, excelling on regular, compute-intensive kernels. However, it also exhibits the greatest performance variability, with its worst-case results often falling below those of the other two architectures. Streaming execution is particularly effective for memory-intensive and partially regular kernels, whereas multicore execution is better suited to irregular, control-dominated workloads and requires the least assembly-level programming effort. These findings provide practical guidance for the hardware–software co-design of heterogeneous architectures that combine complementary execution paradigms.

\printcredits

\bibliographystyle{elsarticle-num}

\bibliography{cas-refs}



\end{document}